\documentclass[prd,aps]{revtex4} 
\usepackage{graphicx}

\begin{document}
\title{Quantum cosmic censor: gravitation makes reality undecidable}

\author{Rodolfo Gambini}
\address{Instituto de F\'{\i}sica, Facultad de Ciencias, 
Universidad
de la Rep\'ublica, Igu\'a 4225, CP 11400 Montevideo, Uruguay\\
rgambini@fisica.edu.uy}
\author{Jorge Pullin}
\address{Department of Physics and Astronomy, 
Louisiana State University, Baton Rouge,
LA 70803-4001\\pullin@lsu.edu}

\date{March 11th 2008}

\begin{abstract}
  When one takes into account gravitation, the measurement of space
  and time cannot be carried out with infinite accuracy.  When quantum
  mechanics is reformulated taking into account this lack of accuracy,
  the resolution of the measurement problem can be implemented via
  decoherence without the usual pitfalls. The resulting theory has the
  same physical predictions of quantum mechanics with a reduction
  postulate, but is radically different, with the quantum states
  evolving unitarily in terms of the underlying variables. Gravitation
  therefore makes this worrisome situation, potentially leading to two
  completely different views of reality, irrelevant from an empirical
  point of view. It may however be highly relevant from a
  philosophical point of view.
\end{abstract}

\maketitle
\eject

This essay links together the findings of several recent papers by the
authors (see \cite{obregon} for a pedagogical review). These papers
draw upon the following well known observations. First of all, it is
well known that in quantum mechanics one needs to expend energy in
order to achieve accurate measurements \cite{saleckerwigner}. On the
other hand, gravity puts fundamental limits on how much energy can be
concentrated in a measuring device before it turns into a black hole.
Coupling together these two observations, one concludes that there
exist fundamental limits, imposed by quantum mechanics and gravity, on
how accurately we can measure distances and time \cite{karolyhazy,ng}.

On the other hand, quantum mechanics is usually formulated on a
space-time manifold that one assumes can be measured with arbitrary
accuracy. Acknowledging that real clocks and rods that one may use to
measure space-time are not arbitrarily accurate requires reformulating
the theory in terms of such clocks and rods \cite{obregon}. It is not
too surprising that in the resulting picture one does not have a
unitary evolution: although the underlying theory is unitary, our
clocks and rods are not accurate enough to give a depiction of
evolution that appears unitary. A detailed calculation shows that
quantum states described in terms of a realistic clock variable $T$
lose coherence as,
\begin{equation}
\rho(T)_{nm}=\rho(0)_{nm}\exp\left(i\omega_{nm} T\right) 
\left(-T_{\rm Planck}^{4/3} T^{2/3} \omega_{nm}^2\right)
\end{equation}
where $\rho(T)_{nm}$ is the density matrix element in an energy
pointer eigenbasis, $\omega_{nm}$ is the Bohr frequency associated
with states $n$ and $m$ in the basis and $T_{\rm Planck}\sim 10^{-44}s$
is Planck's time. The effect is too small to be observed with 
current technologies, but might be  within the reach of technologies of the
relatively near future \cite{SiJa}.

Although experimentally not interesting today, this effect has
important conceptual implications, in particular for the {\em
measurement problem in quantum mechanics} .  The
latter refers to the fact that the system being measured abruptly
falls into an eigenstate right after a measurement has been
performed. This is usually referred to as the reduction process. The
conceptual problem is: how can one explain this abrupt change of
state?  One possible explanation is by saying that there exists an
interaction between the system being measured and the
environment. This interaction selects a preferred basis, i.e., a
particular set of quasi-classical states that commute with the
Hamiltonian governing that interaction.  Decoherence quickly damps
superpositions between the preferred states when only the system is
considered (i.e., when the environment is neglected), and only
classical, well-defined properties appear left to an observer.

The above explanation of the measurement problem encounters some
difficulties. Although we cannot do justice to the full extent of the
problem and its associated vast literature in the confines of this
essay, our claim is that the fact that pure states evolve naturally
into mixed states contributes to surmounting at least some of those
obstacles \cite{measurement}. The usual Hamiltonians describing the
interaction between the system and the environment lead to off
diagonal terms in the density matrix that are oscillatory functions of
time. Typical environments contain a much larger number of degrees of
freedom than the systems being measured.  Therefore the common period
of oscillation for the off diagonal terms to recover nonvanishing
values is very large, often larger than the age of the universe. In
practical terms, the off diagonal elements of the density matrix can
be considered to be zero. But conceptually, there is the remote
possibility that quantum coherence would ``revive''. The fundamental
loss of coherence due to our inability to measure space-time has the
attractive feature of killing off the possibility of ``revivals'' in a
fundamental, inescapable way. If one attempted to ``wait longer'' to
see the effects of revivals, the effect we introduce just becomes
larger.  Therefore off-diagonal terms will never see their initial
values restored, no matter how long one waits. Other ways of
distinguishing between a unitary evolution and the state reduction, as
the ones proposed by d'Espagnat (\cite{despagnat} p. 181) do not appear
possible to be realized due to this process either. We therefore claim that
quantum mechanics formulated with real clocks and rods can, thanks to
fundamental decoherence, does not need a reduction postulate.

This leads us to the main point of the essay: undecidability.  One
cannot decide whether the physical world does in fact contain
reduction of quantum states or not. In fact, it could even be
conceivable that sometimes there might be reduction, sometimes not,
and we do not have reasons to expect one or the other in a given
instance. There is no way to tell between usual quantum mechanics with
a reduction process, on the one hand, and quantum mechanics with real
clocks and decoherence instead of the reduction process. The
difference between these two views of nature is enormous. In one of
them quantum states are given ``once and for all'' as initial
conditions. The evolution is unitary, and what we perceive as loss of
unitarity is due to our inability to access the underlying variables
of the theory, due to gravitational limitations. In the other view,
quantum states are continually evolving due to reduction processes.

In philosophy there are different attitudes that have been taken
towards the physical laws of nature (see for instance
\cite{stanford}).  One of them is the ``regularity theory''; in it,
the laws of physics are statements about uniformities or regularities
of the world and therefore are just ``convenient descriptions'' of the
world. The laws of physics are dictated by a preexisting world and are
a representation of our ability to describe the world. Another point
of view is the ``necessitarian theory'', which states that laws of
nature are ``principles'' which govern the natural phenomena, that is,
the world ``obeys'' the laws of nature. The laws are the cornerstone
of the physical world and nothing exists without a law. The presence of the
undecidability we point out suggests strongly that the ``regularity
theory'' point of view is more satisfactory.

If one takes seriously the regularity point of view one can ponder
about the nature of reality. Does the physical world have a reduction
process, does it not, or does it depend on the case?  In the case in
which there is no reduction process, in the Heisenberg picture the
state of a system is given and eternal. If there is a reduction
process the state changes every time there is an event resulting in a
measurement. The third possibility, which is suggested by the
undecidability, is that the system may choose between behaving as if
there is a reduction process or not.  That
is, after the observation of the event either the system simply
behaves as if it were part of the universe and its state were that of
the universe or if as its state would be given by the reduction
postulate. In the first case the system would keep its entanglement
with the rest of the universe (i.e. the environment), in the second it
will lose its entanglement. This free act of the system will not imply
any violation whatsoever of the laws of physics.

If one adopts what is probably the most attractive view that considers
that the universe always evolves unitarily and therefore quantum
states are determined once and for all no matter what is the chosen
behavior of the subsystem under observation one needs to face the
problem of when do events happen in such a framework. Our point of
view is that an event occurs when the experimental distinction between
coexisting or exclusionary alternatives becomes undecidable, since in
that instant the predictions of the laws of physics are not altered by
the possible reduction of the state of the system associated with the
information acquired when the event takes place.

This situation is unusual in physics: there have never been two theories
that describe reality in rather starkly different ways and nevertheless
have the same experimental predictions. The fact that gravity seems to
be cornering us into this situation is perhaps the ultimate expression
of a sort of ``generalized  cosmic censorship'' in the sense that there
will be aspects of nature, when gravity is taken into account, that
we just will not be able to probe. In this case, the basic structure of
quantum theory itself!

This work was supported in part by grants NSF-PHY0650715,
and by funds of the Horace
C. Hearne Jr. Institute for Theoretical Physics, FQXi, PEDECIBA (Uruguay)
and CCT-LSU.

\end{document}